# ProteinNet: a standardized data set for machine learning of protein structure

Mohammed AlQuraishi[1,*]

[1]Department of Systems Biology, Harvard Medical School, 200 Longwood Ave., Boston, MA, USA

*To whom correspondence should be addressed.

**Abstract**

**Motivation:** Rapid progress in deep learning has spurred its application to bioinformatics problems including protein structure prediction and design. In classic machine learning problems like computer vision, progress has been driven by standardized data sets that facilitate fair assessment of new methods and lower the barrier to entry for non-domain experts. While data sets of protein sequence and structure exist, they lack certain components critical for machine learning, including high-quality multiple sequence alignments and insulated training / validation splits that account for deep but only weakly detectable homology across protein space.

**Results:** We created the ProteinNet series of data sets to provide a standardized mechanism for training and assessing data-driven models of protein sequence-structure relationships. ProteinNet integrates sequence, structure, and evolutionary information in programmatically accessible file formats tailored for machine learning frameworks. Multiple sequence alignments of all structurally characterized proteins were created using substantial high-performance computing resources. Standardized data splits were also generated to emulate the difficulty of past CASP (Critical Assessment of protein Structure Prediction) experiments by resetting protein sequence and structure space to the historical states that preceded six prior CASPs. Utilizing sensitive evolution-based distance metrics to segregate distantly related proteins, we have additionally created validation sets distinct from the official CASP sets that faithfully mimic their difficulty. ProteinNet thus represents a comprehensive and accessible resource for training and assessing machine learned models of protein structure.

**Availability:** Data sets and associated TensorFlow-based parser are available for download at https://github.com/aqlaboratory/proteinnet.

**Contact:** alquraishi@hms.harvard.edu

## 1 Introduction

Deep learning has revolutionized many areas of computer science including computer vision, natural language processing, and speech recognition (LeCun *et al.*, 2015), and is now being widely applied to bioinformatic problems ranging from clinical image classification (Ting *et al.*, 2018) to prediction of protein-DNA binding (Ching Travers *et al.*, 2018). A major driver of the success of deep learning has been the availability of standardized data sets such as ImageNet (Russakovsky *et al.*, 2015), which address three key needs: (i) fair apples-to-apples comparisons with existing algorithms, providing a reference point for the state of the art via a universal benchmark, (ii) at will assessment so that methods can be tried and tested rapidly and new results reported immediately—this has led to weekly publication of new machine learning algorithms—and (iii) access to pre-formatted data with the necessary inputs and outputs for supervised learning. While some bioinformatic applications enjoy this level of standardization (Guinney and Saez-Rodriguez, 2018), the central problem of protein structure prediction remains one without a standardized data set and benchmark. Availability of such a data set can spur new algorithmic developments in protein bioinformatics and lower the barrier to entry for researchers from the broader machine learning community.

Addressing the above needs for protein structure prediction necessitates a data set with several key features. First, sequence and structure data must be provided in a form readily usable by machine learning frameworks, standardizing the treatment of structural pathologies such as missing residues and fragments and non-contiguous polypeptide chains. Second, multiple sequence alignments (MSAs) comprised of evolutionarily related proteins for every structure should be made available, given the central role that evolutionary information plays in modern protein structure prediction (de Oliveira and Deane, 2017). This is especially important as the generation of MSAs can be computationally demanding. Third, standardized training / validation / test splits that partition the data into subsets for fitting model parameters (training set), fitting model hyperparameters (validation set), and model assessment (test set) are needed to ensure consistency when training and assessing different learning algorithms (Goodfellow *et al.*, 2016). Creating such splits can be straightforward for machine learning tasks involving images or speech, as data points from these modalities can be reasonably approximated as independent and identically distributed (IID). Natural protein sequences are far from IID however due to their underlying evolutionary relationships, a problem further exacerbated by the discrete nature of these sequences which can result in similar proteins having nearly identical numerical representations (this problem is avoided by e.g. images, as even small changes in lighting or viewing angle result in entirely different pixel-level representations). Consequently careful treatment of data splitting is required to ensure meaningful separation between subsets. Finally, multiple test objectives should ideally be provided to enable nuanced assessment of new methods, for example by testing varying levels of generalization capacity.

While a variety of protein structure databases do exist, none satisfy all the above requirements. Repositories such as the Protein Data Bank (PDB) (Bernstein *et al.*, 1977) provide raw protein structures, but require post-processing before they are usable by machine learning frameworks. Processed data sets such as CulledPDB (Wang and Dunbrack, 2003) provide a more standardized preparation of protein structures, but lack evolutionary data such as MSAs. In fact, to our knowledge there is currently no public resource for high-quality MSAs suitable for protein structure prediction. One MSA repository does exist (Joosten *et al.*, 2011), but it appears out of date and is unsuitable for applications requiring deep homology searches (Rost, 1999). The substantial computational cost associated with generating MSAs may explain this surprising absence.

With respect to standardized training / validation / test splits, the closest existing analogues are the biennial Critical Assessment of protein Structure Prediction (CASP) (Moult John *et al.*, 2018) and the continually running Continuous Automated Model EvaluatiOn (CAMEO) (Haas *et al.*, 2018). Both of these ongoing experiments provide an invaluable service for assessing prediction methods in a blind fashion, by presenting predictors with protein sequences whose structure has been solved but not yet made publicly available. Nonetheless, these experiments serve a different purpose from a standardized data split. CASP occurs once every two years, making it too infrequent for rapidly developing fields like machine learning. And while CASP can be thought to provide a training / test split based on the data available on the starting day of a given CASP assessment, it does not provide a validation set. Effective validation sets must mimic the generalization challenge presented to a trained model by the test set, by mirroring the distributional shift in data between the training and test sets; in effect, acting as a proxy for the test set. This is challenging to do for CASP proteins as they often contain novel protein folds occupying the twilight zone of sequence homology relative to PDB proteins (<30% sequence identity (Khor *et al.*, 2015)). Creating a matching validation set is thus non-trivial owing to the difficulty of detecting weak homology (Habermann, 2016; Chen *et al.*, 2018).

Unlike CASP, CAMEO is continually running and thus can be used for assessment at any time. However, by virtue of its dynamic nature, CAMEO is difficult to use for apples-to-apples comparisons with an existing method unless both methods are participating simultaneously. CAMEO also focuses on proteins with known folds, making it less suitable for testing generalization to unknown parts of protein fold space.

To address these challenges and provide a community resource that promotes the application of machine learning to protein structure, we created ProteinNet. ProteinNet provides pre-formatted input / output records comprising protein sequences, high-quality MSAs, and secondary and tertiary structures, as well as standardized data splits, including validation sets that emulate the generalization challenges presented by CASP proteins.

## 2 Methods

### 2.1 Design and approach

ProteinNet's design philosophy is simple: piggyback on the CASP series of assessments to create a corresponding series of data sets in which the test set is comprised of all structures released in a given CASP, and the training set is comprised of all protein structures and sequences (for building MSAs) publicly available prior to the start date of that CASP. A subset of the training data is set aside to create multiple validation sets at different sequence identity thresholds (relative to the training set), including <10% to test generalizability to new protein folds comparable in difficulty to those encountered in CASP. Each ProteinNet data set effectively reverts the historical record to mimic the conditions of a prior CASP. We use CASP 7 through 12 (dating back to 2006) to create the corresponding ProteinNet 7 through 12. Our approach has three desirable properties.

First, by utilizing CASP structures for the test set, we leverage an objective third party's (the CASP organizers') selection of structures that meaningfully differ from the publicly accessible universe of PDB structures at a given moment in time. In particular, CASP organizers place prediction targets in two categories: template-based modeling (TBM) for proteins with clear structural homology to PDB entries, and free modeling (FM) for proteins containing novel folds unseen or difficult to detect in the PDB. This delineation provides an independent measure of difficulty useful for assessing models' ability to generalize to unseen parts of fold space. (CASP organizers occasionally include a third category, "TBM / FM" or "TBM hard", for structures of medium difficulty.)

Second, by utilizing multiple validation sets with varying levels of sequence identity, ProteinNet provides proxies for both TBM (20% - 40% seq. id.) and FM (<10% seq. id.) CASP proteins. This enables optimization of model hyperparameters through monitoring of model generalization to proteins similar in difficulty to CASP TBM or FM proteins, potentiating the development of models focused exclusively on novel or known fold prediction. We note that this distinct from merely having separate TBM and FM test proteins (first property), as test sets are only used for final model assessment and are thus unsuitable for hyperparameter optimization (the purpose of validation sets).

Third, by virtue of being the standard for assessing structure prediction methods, CASP enjoys the participation of all major predictors. It thereby provides a record of the accuracy of current and prior methods given available data at assessment time. Crucially, new methods trained and tested on ProteinNet demonstrate their performance on the same data splits as CASP-assessed methods, making them immediately comparable to state of the art methods on current and prior CASPs. This circumvents the catch-22 problem facing new benchmarks by providing immediate value to ProteinNet-trained models. Comparisons using older CASPs provide assessments with varying amounts of data, stressing algorithms in data rich and data poor regimes, a useful property when controlling for algorithmic vs. data-driven improvements, particularly in co-evolution-based methods.

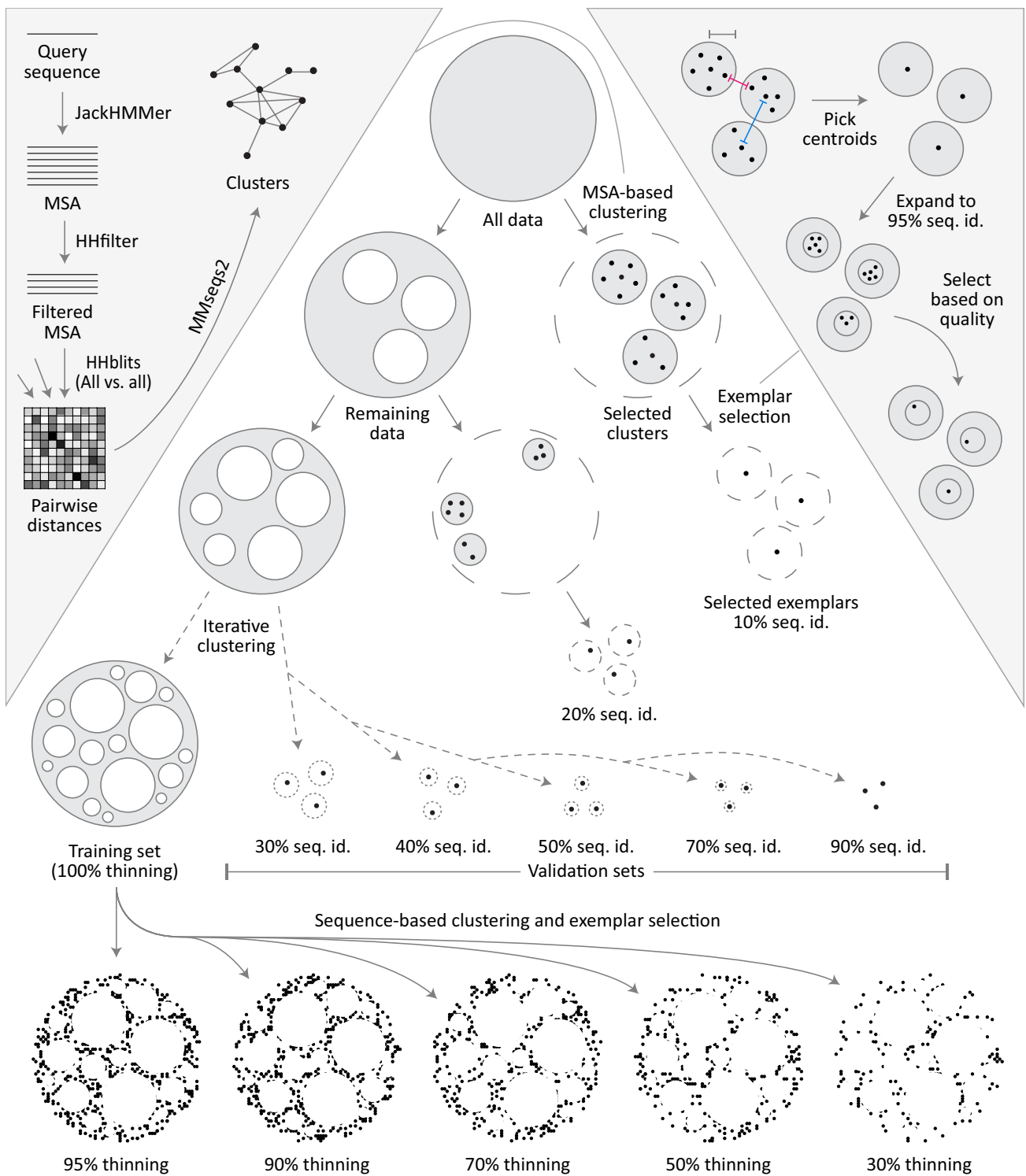

**Fig. 1. ProteinNets construction pipeline.** For each ProteinNet, all proteins with PDB structures available prior to the start of its corresponding CASP ("All data", top circle) are clustered using an MSA-based clustering technique (left inset) to yield large clusters where intra-cluster sequence identity is as low as 10%. One exemplar from each cluster is then selected (right inset) to yield the 10% seq. id. validation set. This process is iteratively repeated, by reclustering the data remaining outside of all initial clusters to yield validation sets of higher sequence identity (20% - 90%). Once the final validation set is extracted, all remaining data is used to form the training set. Based on this set ("100% thinning"), filtered training sets are created at lower sequence identity thresholds to provide coarser sampling of sequence space. **Left inset:** Each protein sequence is queried against a large sequence database (filtered to only include sequences publicly available prior to the beginning of the corresponding CASP) using JackHMMer to create an MSA that is subsequently filtered to 90% seq. id. HHblits is then used to perform an all-against-all sequence alignment of MSAs. Finally, alignment distances are fed to MMseqs2 to cluster their corresponding sequences. **Right inset:** The center-most protein of each cluster is chosen to ensure that the desired sequence identity constraints are satisfied, as proteins near cluster boundaries may be closer than the pre-specified radius of each cluster (pink vs. gray measuring tapes), while the distances between cluster centroids must satisfy the sequence identity constraints (blue measuring tape). The centroids are then used to form tight clusters of 95% seq. id. that are intersected with the original clusters to yield candidate exemplars ranked by multiple quality metrics (see main text). The top-ranked candidate is picked as the exemplar protein of each cluster.

**Table 1.** ProteinNet cutoff dates for data inclusion, based on the start of CASP experiments, and number of sequences and structures in each set.

| Data set | Cutoff date | Structures* | Sequences* |
|---|---|---|---|
| ProteinNet 7 | 2006/5/1 | 34,557 | 4,817,827 |
| ProteinNet 8 | 2008/5/5 | 48,087 | 15,756,117 |
| ProteinNet 9 | 2010/5/3 | 60,350 | 24,688,095 |
| ProteinNet 10 | 2012/5/1 | 73,116 | 63,477,198 |
| ProteinNet 11 | 2014/5/1 | 87,573 | 173,908,140 |
| ProteinNet 12 | 2016/5/1 | 104,059 | 332,283,871 |

* Non-redundant

### 2.2 Structures and sequences

All current PDB structures were downloaded using the mmCIF file format (Westbrook and Fitzgerald, 2005) then filtered by public release date so that ProteinNet training and validation sets only include entries publicly available prior to the start of their corresponding CASP assessment (Table 1). We exclude structures containing less than two residues or where >90% of residues were not resolved, but otherwise retain virtually the entirety of the PDB. Mask records are generated for each structure to indicate which residues or fragments, if any, are missing, to facilitate processing by machine learning algorithms (e.g. by using a loss function that ignores missing residues). Multiple logical chains (in the same mmCIF file) that correspond to a single physical polypeptide chain are combined (with missing fragments noted), while physically distinct polypeptide chains are treated as separate structures. For chains with multiple models, only the first one is kept.

Sequences are derived directly from the mmCIF files. In instances where an amino acid is chemically modified or its identity is unknown, the most probable residue in its position-specific scoring matrix (PSSM) (Stormo, 2000) is substituted (see next section for how PSSMs are derived). If a PSSM contains more than three adjacent residues with zero information content then its corresponding sequence / structure is dropped, as we have found this to be a strong indicator that the sequence cannot be faithfully resolved.

In addition to full length PDB structures and sequences, single domain entries are created by extracting domain boundaries from ASTRAL (Fox *et al.*, 2014) to enable training of both single and multi-domain models.

### 2.3 Multiple sequence alignments

Sequence databases for deriving MSAs were created by combining all protein sequences in UniParc (UniProt Consortium, 2018) with metagenomic sequences from the Joint Genome Institute (Ovchinnikov *et al.*, 2017) and filtering to include only sequences available prior to CASP start dates (Table 1). JackHMMER (Eddy, 2011) was then used to construct MSAs for every structure by searching the appropriate sequence database. Different MSAs were derived for the same structure if it occurred in multiple ProteinNets. JackHMMER was run with an e-value of 1e-10 (domain and full length) and five iterations. A fixed database size of 1e8 (option -Z) was used to ensure constant evolutionary distance when deriving MSAs (similar to using bit scores). Only perfectly redundant sequences (100% seq. id.) were removed from sequence databases to preserve fine- and coarse-grained sequence variation in resulting MSAs.

In addition to raw MSAs, PSSMs were derived using Easel (Potter *et al.*, 2018) in a weighted fashion so that similar sequences collectively contributed less to PSSM probabilities than diverse sequences. Henikoff position-based weights (option -p) were used for this purpose.

### 2.4 Data splits and thinning

For each CASP cutoff date, we partition the full complement of (pre-CASP) structures and their associated MSAs into one training set and multiple validation sets, all non-overlapping (Fig. 1). Partitioning is done iteratively, by first clustering sequences at the 10% sequence identity level, randomly extracting 32 clusters, and then reclustering the remaining sequences at the next sequence identity level. Seven thresholds are used (10%, 20%, 30%, 40%, 50%, 70%, 90%) resulting in seven validation sets each comprising 32 clusters. While the selection of clusters is random, clusters larger than 100 members are not considered to minimize data loss. One representative exemplar is then selected from each cluster and the remaining cluster members are removed entirely (exemplar selection criteria is described at the end). Structures that remain outside of all validation clusters comprise the training set.

Obtaining coherent clusters at <20% sequence identity is difficult due to weak homology between individual sequences. To overcome this we perform comparisons using the previously derived MSAs instead of using individual sequences, as they provide greater sensitivity by incorporating evolutionary information (left inset in Fig. 1). First, sequences within MSAs are redundancy filtered to 90% seq. id. using HHsuite (Söding, 2005) to lower the computational load. We then carry out an all-against-all MSA-to-MSA comparison using HHblits (Remmert *et al.*, 2012) with one iteration and local alignment (option -loc). HHblits is necessary for this step as JackHMMER is unable to perform MSA-to-MSA searches, but the MSAs used are the original, JackHMMER-derived ones. Based on the HHblits alignment scores, we cluster MSAs using MMseqs2 (Steinegger and Söding, 2017, 2) with the sought sequence identity threshold, an e-value threshold of 0.001, and clustering mode 1, which constructs a graph covering all sequences then finds remote homologs using transitive connections. We do not impose a minimal coverage requirement on sequence hits; this overestimates sequence identity, as short proteins can match subparts of longer ones. We use this approach to be maximally conservative in our construction of validation sets, to safeguard against accidental information leakage between training and validation sets.

Training sets are further processed to generate overlapping subsets that vary in sequence redundancy (at 30%, 50%, 70%, 90%, 95%, and 100% seq. id.) which we call "thinnings". For every thinning except 100% we cluster the training set by applying MMseqs2 directly to individual sequences (no MSAs) with the sought sequence identity threshold and a sequence coverage requirement of 80%. This requirement ensures that individual domains are not grouped with multi-domain proteins that contain them, as some models may seek to leverage single domain information. We do not utilize a coverage requirement for the validation set to prevent information leakage, but it is not a concern for the training set. For the 100% thinning every set of identical sequences is used to form a cluster. After clusters are formed, a single exemplar is selected from each, and all remaining cluster members are removed.

We use the same exemplar selection criteria for validation and training sets. As a rule, we avoid selecting exemplars near cluster boundaries, as two boundary sequences in different clusters may be closer in sequence space than the sought sequence identity threshold (right inset in Fig. 1). To ensure this we always pick exemplars near the cluster center. By default, MMseqs2 returns an exemplar which is centermost in the cluster without incorporating other, potentially useful criteria such as structure quality. To incorporate these criteria we use the MMseqs2 exemplar as bait to form a new cluster of sequences that are 95% identical to the exemplar and cover 90% of its length, yielding a tight cluster that is highly sequence-similar but with potentially better structural characteristics. From the intersection of the original MMseqs2-derived cluster and the new one, we then pick the structure that optimizes the following criteria, in order: structure quality (defined as 1 / resolution - R-value, the same criterion used by the PDB), date of release (newer is better), and length (longer is better).

### 2.5 File formats and availability

All sequences, structures, MSAs, and PSSMs have been made available for download individually in standard file formats. In addition, ProteinNet records integrating sequence, structure (secondary and tertiary), and PSSMs in a unified format are available as human-readable text files and as binary TensorFlow records (Abadi *et al.*, 2016). We provide Python code for parsing these records directly into TensorFlow to facilitate their use in machine learning applications.

## 3 Results

We applied the ProteinNet construction pipeline to CASP 7 through 12, resulting in six data sets ranging in size from 34,557 to 104,059 structures (Table 1). We observe a generally linear increase in the number of training structures, across all thinnings, over this time period (Fig. 2A), consistent with the PDB's sequence bias remaining constant. The growth in sequence data on the other hand is exponential (Table 1), much of which driven by metagenomic databases comprised largely of prokaryotic genes. Since unknown prokaryotic genes are less likely to be crystallized, they are not well presented among CASP targets (Ovchinnikov *et al.*, 2017). Nonetheless, the growth in sequence databases has resulted in higher quality MSAs in later CASPs, as measured by the number of sequences in alignments (Fig. 4). The overall number of CASP test structures is roughly constant, although the proportion of FM targets has increased (Fig. 3), likely reflecting the end of the Structural Genomics Initiative (Chandonia and Brenner, 2006) which crystallized a large number of proteins of known folds.

Examining sequence length, we observe that CASP structures have on average grown in length (Fig. 3), and similarly for the PDB (Fig. 2B,C), although the vast majority of structures remain shorter than 1,000 residues. This trend is likely to accelerate with increased use of CryoEM (Callaway, 2015) methods which have made multi-domain proteins more amenable to structural characterization.

To assess the suitability of ProteinNet validation sets to serve as proxies for CASP targets, we computed the distance, measured by sequence identity, of every entry in the ProteinNet validation and test sets to its closest entry in the training set. Because sequence identity is difficult to detect in low homology regions (<30% seq. id.), we first performed an all-against-all alignment using MSA-MSA searches, similar to our pipeline for clustering, and then computed sequence identity using the resulting matches. We used an e-value threshold of 0.001 to ensure genuine hits, but otherwise imposed no additional constraints. Fig. 5 summarizes the results. As expected, FM targets across all CASPs show no detectable sequence homology to their corresponding ProteinNet training sets. Importantly, the <10% seq. id. validation sets of all ProteinNets show no detectable homology to the training sets either, indicating that they can act as proxies of CASP FM targets. TBM targets roughly exhibit between 10% and 30% seq. id. to the ProteinNet training sets, similar to the <20%, <30%, and <40% seq. id. validation sets, confirming that they can as proxies of CASP TBM targets. We conclude that the appropriate ProteinNet validation set can be used to optimize models whose goal is to generalize to protein folds similar in difficulty to CASP FM and TBM targets. ProteinNet validation sets with higher sequence identity, i.e. >50%, are potentially useful for optimizing models focused on predicting changes to known protein structures, such as those induced by mutations.

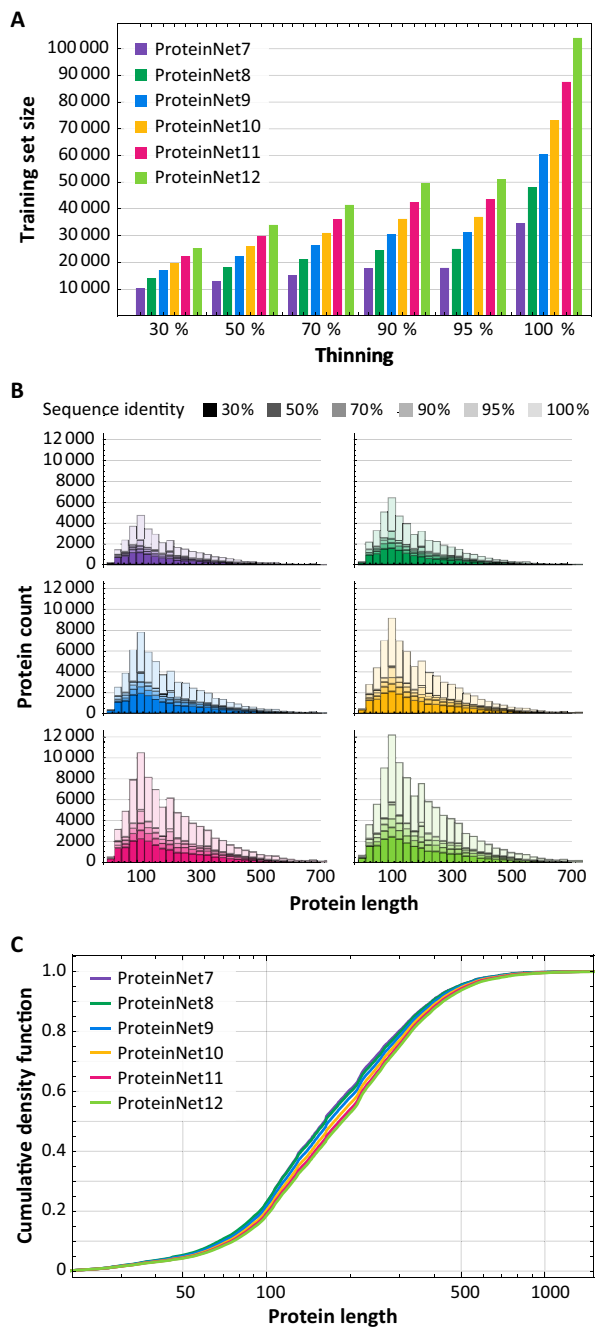

**Fig. 2. Statistics of ProteinNet data sets. (A)** Number of proteins in ProteinNet training sets for different thinnings (30% - 100% seq. id.) **(B)** Protein length distributions for ProteinNet training sets. **(C)** Cumulative density function of protein length distribution for 100% thinnings of ProteinNet training sets.

We next sought to assess how growth in the number of PDB structures changes the difficulty of CASP TBM targets. For every CASP test set, we repeated the previous analysis using older ProteinNet training sets. E.g., for CASP 11, we compared its TBM set against ProteinNet 7 – 11 training sets. Fig. 6 summarizes the results. As expected, earlier ProteinNet training sets show greater distance from the TBM sets, particularly for older CASPs, with a general loss of ~2-3% seq. id. points per CASP (i.e. two years). This type of retroactive analysis may be used to assess an algorithm's sensitivity to the amount of available data, with Fig. 6 providing a characterization of the relative difficulty of different CASP targets when using different ProteinNets for training (raw distance data at the single protein level is available at the ProteinNet repository). We did not perform this analysis for FM targets since even the most up to date ProteinNet training sets (for a given CASP) do not show any detectable homology, thereby precluding older training sets from showing further homology.

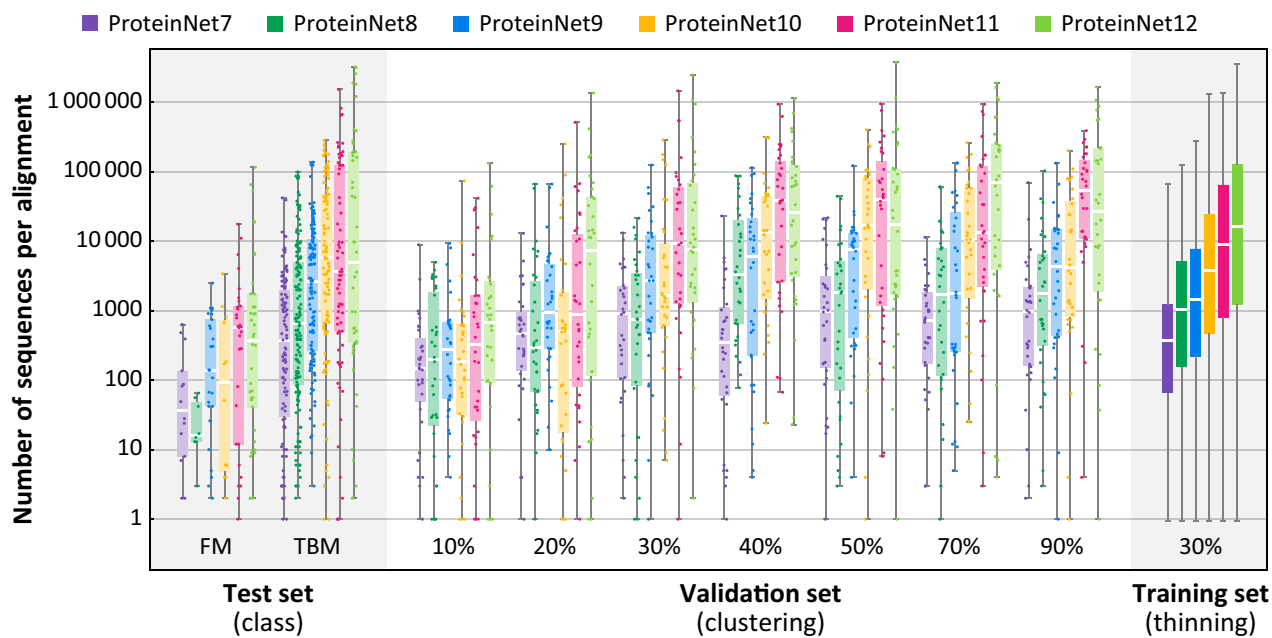

**Fig. 3. Alignment size as a function of ProteinNet subset.** Box and whisker charts depict the distribution of number of sequences per MSA for ProteinNet training (30% thinning), validation, and test sets. Individual data points for training sets are not shown due to their large size.

## 4 Discussion

Standardized data sets have unlocked progress in myriad areas of machine learning, and biological problems are no exception. ProteinNet represents a community resource for bioinformaticists and machine learning researchers who seek to test new algorithms in a manner consistent with state of the art blind assessment. It lowers the barrier to entry for the field by aggregating the relevant data modalities in a single file format, and by eliminating the upfront computational cost required for creating high-quality MSAs. Collectively, the generation of all MSAs and PSSMs in ProteinNet 7 – 12 consumed over 3 million compute hours, a one-time investment whose benefits can now be shared by the entire community of researchers. Perhaps most crucially, ProteinNet provides validation sets that provide a reliable assessment of model generalizability, ensuring that progress can be meaningfully ascertained while training models.

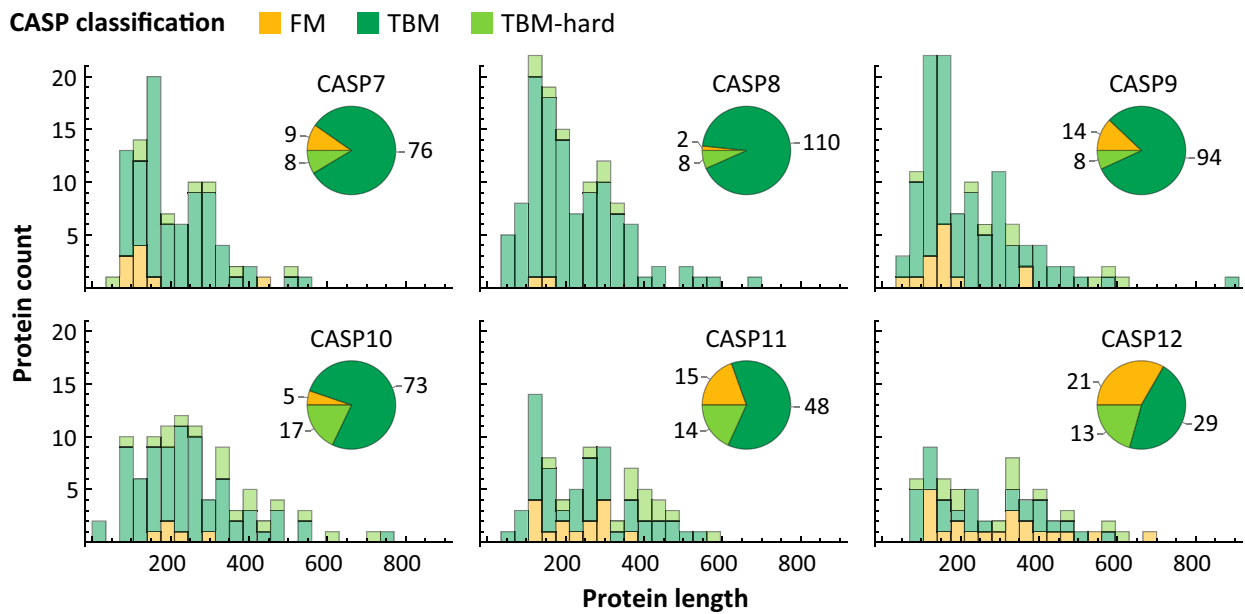

**Fig. 4. Statistics of CASP data sets.** Length distribution of proteins in CASP 7 through 12, broken down by difficulty class. Pie charts show the number of proteins per difficulty class.

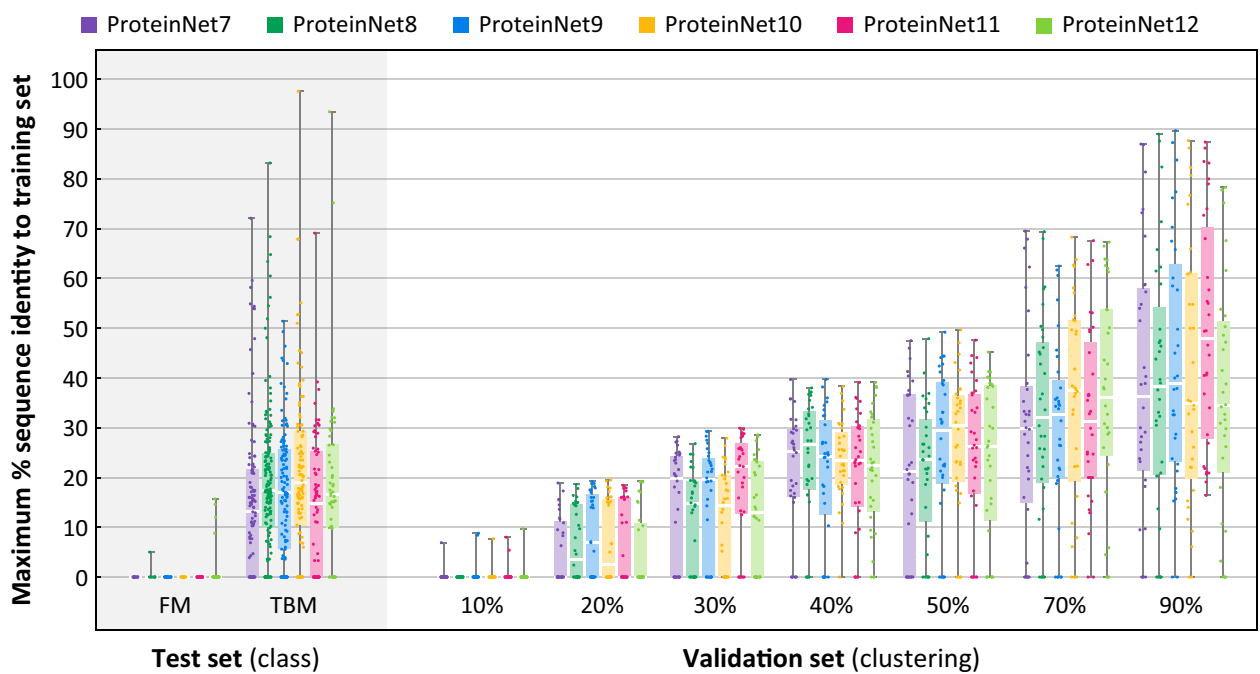

**Fig. 5. Distributions of maximum % sequence identity to training sets.** Box and whisker charts depict the distribution of maximum % sequence identity, with respect to the training set, of all entries in a given ProteinNet validation or test set. The FM test sets and 10% seq. id. validation sets show a median value of 0% seq. id. to the training set.

Beyond protein structure prediction, ProteinNet can serve as a data set for a number of important problems. ProteinNet prescribes no intrinsic preference for which data modalities should serve as inputs and which should serve as outputs. A protein design algorithm can hypothetically be trained by using structures as inputs, and the sequences of their associated MSAs as outputs. Alternatively, an algorithm for predicting the effects of mutant variants can use the sequence and structure of one protein as input, and output the structures of proteins with similar sequences as predictions.

More broadly, the advent of deep learning methods and automatic differentiation frameworks like TensorFlow and PyTorch (Paszke *et al.*, 2017) makes it possible to build bespoke models of biological phenomena.

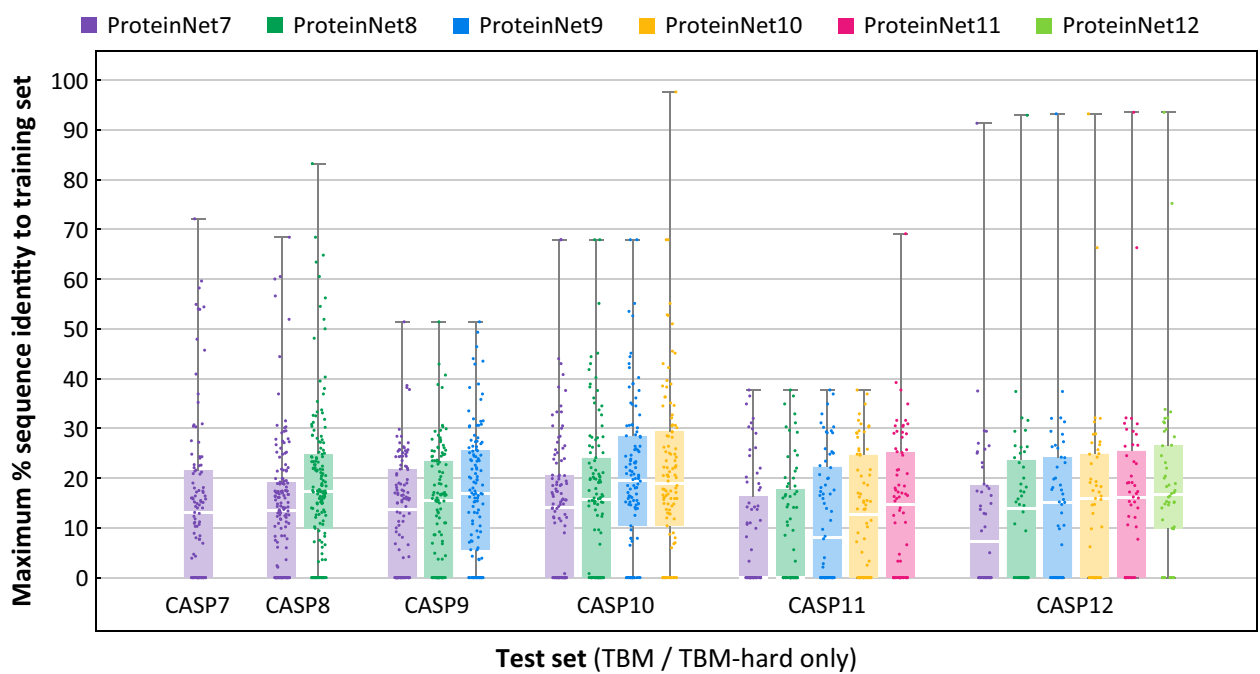

**Fig. 6. Distributions of maximum % sequence identity of CASP entries with respect to prior training sets.** Box and whisker charts depict the distribution of maximum % sequence identity, with respect to a training set, of all TBM / TBM-hard entries in a given ProteinNet test set (CASP set). Comparisons are made for each ProteinNet test set with respect to its corresponding and prior training sets, e.g. for CASP 11 with respect to ProteinNet 7 – 11 training sets. Color indicates training set used.

In the machine learning community, this has spurred the development of so-called multi-task learning problems in which multiple output modalities are simultaneously predicted from a given input, as well as auxiliary losses in which a core objective function is augmented with additional output signals that can help train a more robustly generalizing model. In many gene- or protein-related learning tasks, protein structure is one such broadly useful output signal that can augment a supervised learning problem, e.g. the prediction of the DNA binding affinity of a transcription factor, with information that is proximal to the desired task. ProteinNet should help facilitate such applications, along with the development of end-to-end differentiable models of protein structure that can be directly fused to other learning problems (AlQuraishi, 2018). As the quality of protein structure prediction algorithms continues to improve, we believe that structural information will get increasingly integrated within a wide swath of computational models.

## Acknowledgements

We thank Peter Sorger for his mentorship and support, and Uraib Aboudi for her editorial comments and helpful discussions. We also thank Martin Steinegger and Milot Mirdita for their help with using the HHblits and MMseqs2 packages, Sergey Ovchinnikov for help with metagenomics sequences, Andriy Kryshtafovych for his help with CASP structures, Sean Eddy for his help with using the JackHMMer package, and Raffaele Potami, Amir Karger, and Kristina Holton for their help with using HPC resources at Harvard Medical School.

## Funding

This work has been supported by NIGMS Grant P50GM107618.

*Conflict of Interest:* none declared.